\documentclass[aps,prb,twocolumn,amsmath,amssymb]{revtex4-2}

\usepackage{graphicx}
\usepackage{bm}
\PassOptionsToPackage{hyphens}{url}
\usepackage{hyperref}
\usepackage{xcolor}
\usepackage[normalem]{ulem} 
\usepackage{cancel}

\begin{document}

\title{Stable levitation of ring magnets: A low-cost, 3D-printed experiment on the magnetic force--distance law}

\author{M. Flores}
\email{maicol.sebastian@fcien.edu.uy}
\affiliation{Facultad de Ciencias, Universidad de la Rep\'ublica, Montevideo, Uruguay}

\author{A. Kahrs}
\affiliation{Facultad de Ciencias, Universidad de la Rep\'ublica, Montevideo, Uruguay}

\author{F. Rinderknecht}
\affiliation{Facultad de Ciencias, Universidad de la Rep\'ublica, Montevideo, Uruguay}

\author{Arturo C. Mart\'i}
\affiliation{Facultad de Ciencias, Universidad de la Rep\'ublica, Montevideo, Uruguay}

\date{\today}

\begin{abstract}
Magnetic levitation is one of the most captivating phenomena in physics, combining a striking visual appeal with the rich underlying physics of forces and interactions in equilibrium. Bringing it into undergraduate laboratories in a quantitative way is nevertheless often hindered by the intrinsic instability of static magnetic systems and by the high cost of active control systems.
Here we present a simple, safe, and very low-cost experiment that uses ring-shaped ferrite magnets together with a mechanical stabilization system fabricated by 3D printing in a base--piston configuration. The apparatus lets undergraduate students explore directly and quantitatively the relationship between the magnitude of the magnetic force and the separation distance. We assess the robustness of the device through experimental tests, whose results confirm the inverse dependence of the equilibrium separation on the applied load predicted by the equivalent current-loop model. Furthermore, we cross-check the effective model parameters with a complementary measurement of the axial magnetic field. To support reproducibility, the 3D design files are made available in a public, open-access repository.
\end{abstract}

\maketitle

\section{Introduction}

The study of permanent magnets and their mutual interactions is a fundamental component of the general-physics and electromagnetism curriculum.\cite{gonzalez_ejp_2017, forringer_tpt,arribas2015measurement} Magnetic levitation, in particular, has extraordinary instructional value because of its counterintuitive nature and its ability to spark genuine student interest.\cite{tada_tpt_2026} Beyond its undeniable visual and sensory appeal, however, this setting poses a deep conceptual challenge concerning the notion of force, Newton's laws, and the conditions for static equilibrium.

According to Earnshaw's theorem,\cite{earnshaw} it is impossible to maintain a stable levitation configuration in three-dimensional space using only static permanent magnets.\cite{scott_earnshaw} This restriction forces the design of alternative strategies to achieve stability, which span a broad range of physical principles.\cite{rossing_hull_1991} Among them are gyroscopic stabilization,\cite{simon_ajp_1997} and the use of diamagnetic or superconducting materials.\cite{berry_geim_1997, osorio_2012, giliberti_ejp_2018, meissner_1933, strehlow_ajp_2009} Unfortunately, these solutions usually add mathematical complexity or economic cost that hinders their widespread adoption in undergraduate teaching laboratories.

 The proposal presented here was successfully posed as an experimental problem in the national Physics Olympiads. Such contests create a permanent demand for ingenious, original, and motivating experiments that, at the same time, lend themselves to the low-cost, parallel assembly of many identical kits, so that all participants can work simultaneously under the same conditions. Here, the rise of 3D printing by fused deposition modeling (FDM) offers a route to the rapid, low-cost fabrication of customized laboratory devices with high geometric fidelity,\cite{tomes_ejp_2016, bley_ejp_2021} facilitating the transition toward the visualization of abstract and spatial concepts in science education.\cite{su_ajp_2016, rossi_ajp_2021} Following this philosophy, this article presents a proposal that overcomes the Earnshaw instability through a simple and elegant one-dimensional geometric constraint, using accessible components and 3D printing.\cite{perez_ajp_2019, malmstrom_ejp_2020} Similar low-cost configurations and analytical models have been explored in the broader literature to characterize the interaction of magnets and the corresponding power laws in educational settings \cite{ref_1_13365, ref_2768690, ref_0143_0807}. The system uses two ring-shaped ferrite magnets oriented in opposition. While the lower magnet remains fixed, the upper magnet levitates in a state of mechanical equilibrium in which the magnitude of the magnetic force (which is repulsive in this configuration) exactly balances the total weight of the levitating system. By sequentially adding non-magnetic masses, one can obtain the relationship between the magnitude of this magnetic force and the separation distance, giving undergraduate students a direct, quantitative experience in force analysis.

\section{Equivalent current-loop model}

To model the interaction between the two ring-shaped magnets analytically, we adopt a classical approximation in which each ring magnet of radius $R$ is represented geometrically as an ideal single-turn coil carrying an effective current $I$. This approach relies on the physical equivalence of Amperian surface currents, providing a consistent macroscopic model.\cite{vuckovic_icest, rossing_hull_1991, saslow_tpt_2022} Within this framework, the magnitude of the magnetic force $F_{\text{mag}}$ (which acts repulsively) exerted by the lower magnet on the upper one is given by
\begin{equation}
F_{\text{mag}} = \ln(4)\frac{\mu_{0}\, I^{2} R }{h},
\label{eq:fuerza}
\end{equation}
where $h$ is the separation between the magnets, $R$ is an effective loop radius (taken here as the outer radius of the ring), $I$ is the effective current, and $\mu_{0}~= 4\pi\times10^{-7}~\text{N/A}^2$ is the permeability of free space.\cite{perez_ajp_2019, vuckovic_icest, robertson_ieee} A detailed step-by-step mathematical derivation of Eq.~(1), including the limits of the near-field approximation and the geometric origin of the $\ln(4)$ scaling factor, is explicitly detailed in Appendix~\ref{app:deduction}.

In static equilibrium, the net force vector on the upper magnet must vanish. The magnitude of the magnetic force therefore balances exactly the total weight of the levitating system, which comprises the mass of the upper magnet ($m_{\text{magnet}}$) plus any additional mass ($m_{\text{add}}$) placed on top of it,
\begin{equation}
F_{\text{mag}}= \left(m_{\text{magnet}} + m_{\text{add}}\right) g,
\label{eq:equilibrio}
\end{equation}
where $g$ is the gravitational acceleration. Combining Eqs.~\eqref{eq:fuerza} and \eqref{eq:equilibrio} shows that the equilibrium separation $h$ is inversely proportional to the applied weight.

Consequently, the model predicts a linear relationship between the inverse separation $h^{-1}$ and the magnitude of the equilibrium magnetic force. This prediction allows students not only to verify the physical law but also to determine experimentally an intrinsic parameter of the magnet, namely its effective current $I$ or its magnetic dipole moment $\mu$.

Alternatively, the axial magnetic field $B(z)$ generated by a single ring magnet at a distance $z$ along its symmetry axis can be modeled by considering the magnet as an equivalent thin current-loop. Under this representation, the on-axis magnetic field strength can be written as a function of the magnetic dipole moment $\mu$ and a purely geometric factor $g(z)$ according to \cite{griffiths_ed}:
\begin{equation}
B(z) = \mu\, g(z) = \mu \left[ \frac{\mu_0}{2\pi} \frac{1}{(z^2 + R^2)^{3/2}} \right].
\label{eq:Bz_model}
\end{equation}
This alternative expression provides an independent, first-principles-based method to characterize the magnet’s parameters through a fit.

\section{Experimental method}

\begin{figure}[h!]
\centering
\includegraphics[width=0.3\textwidth]{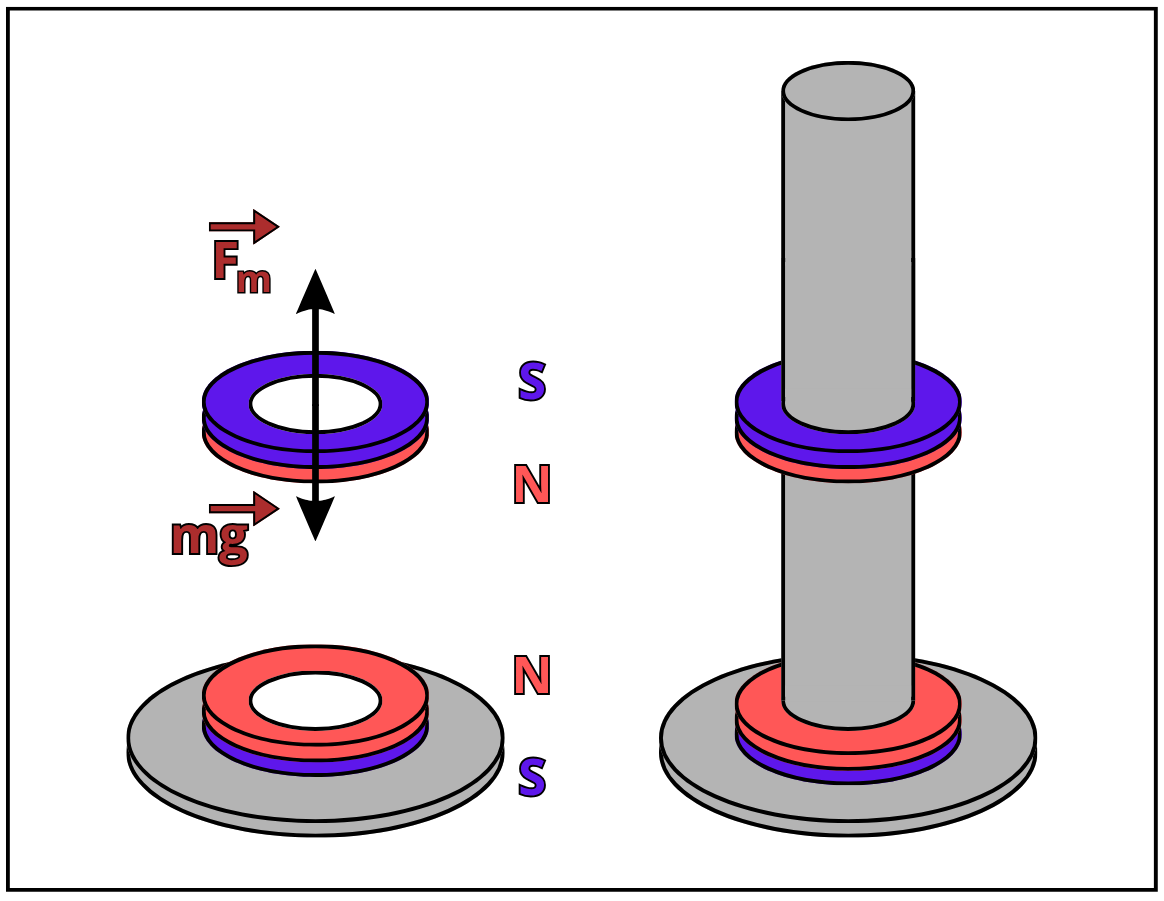}
\caption{Illustration of magnetic levitation. On the left, two ring-shaped magnets face each other, one above the other. On the right, the same arrangement is shown with a cylindrical guide that provides stability to the levitating magnet.}
\label{fig:ilustracion_levitacion}
\end{figure}

The experimental setup was designed for simplicity and reproducibility. To counteract the natural tendency of the upper magnet to flip and stick laterally to the lower one---as dictated by the Earnshaw instability---we designed a coaxial cylindrical guide. The apparatus utilizes two identical ferrite ring magnets with an internal diameter of $17~\text{mm}$, an external diameter of $32~\text{mm}$, and an axial thickness of $6~\text{mm}$. The conceptual principle of the assembly and its transition to a stable solution are illustrated in Fig.~\ref{fig:ilustracion_levitacion}.

\begin{figure}[h!]
\centering
\includegraphics[width=0.34\textwidth]{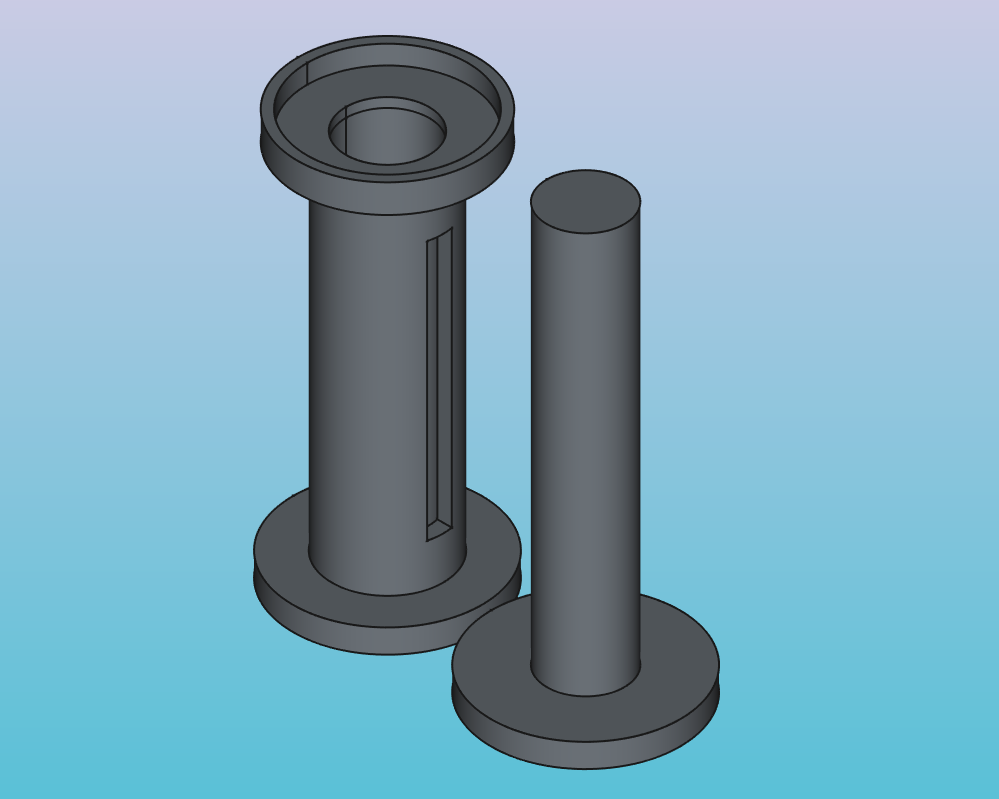}
\caption{Experimental design produced for 3D printing. On the left is the ``base'' part and on the right the ``piston'' part.}
\label{fig:diseno_3d}
\end{figure}

To realize the cylindrical guide while minimizing friction and ensuring perfect concentric alignment, we modeled and 3D-printed a coupled base--piston device using ABS (acrylonitrile butadiene styrene), a non-magnetic material that does not interfere with the field lines. The geometric details of the printed parts are shown in Fig.~\ref{fig:diseno_3d}. To facilitate reproduction of the setup in other educational settings, the 3D models are provided as STL files in a public, open-access repository (see Sec.~\ref{sec:availability}).

\begin{figure}[h!]
\centering
\includegraphics[width=0.48\textwidth]{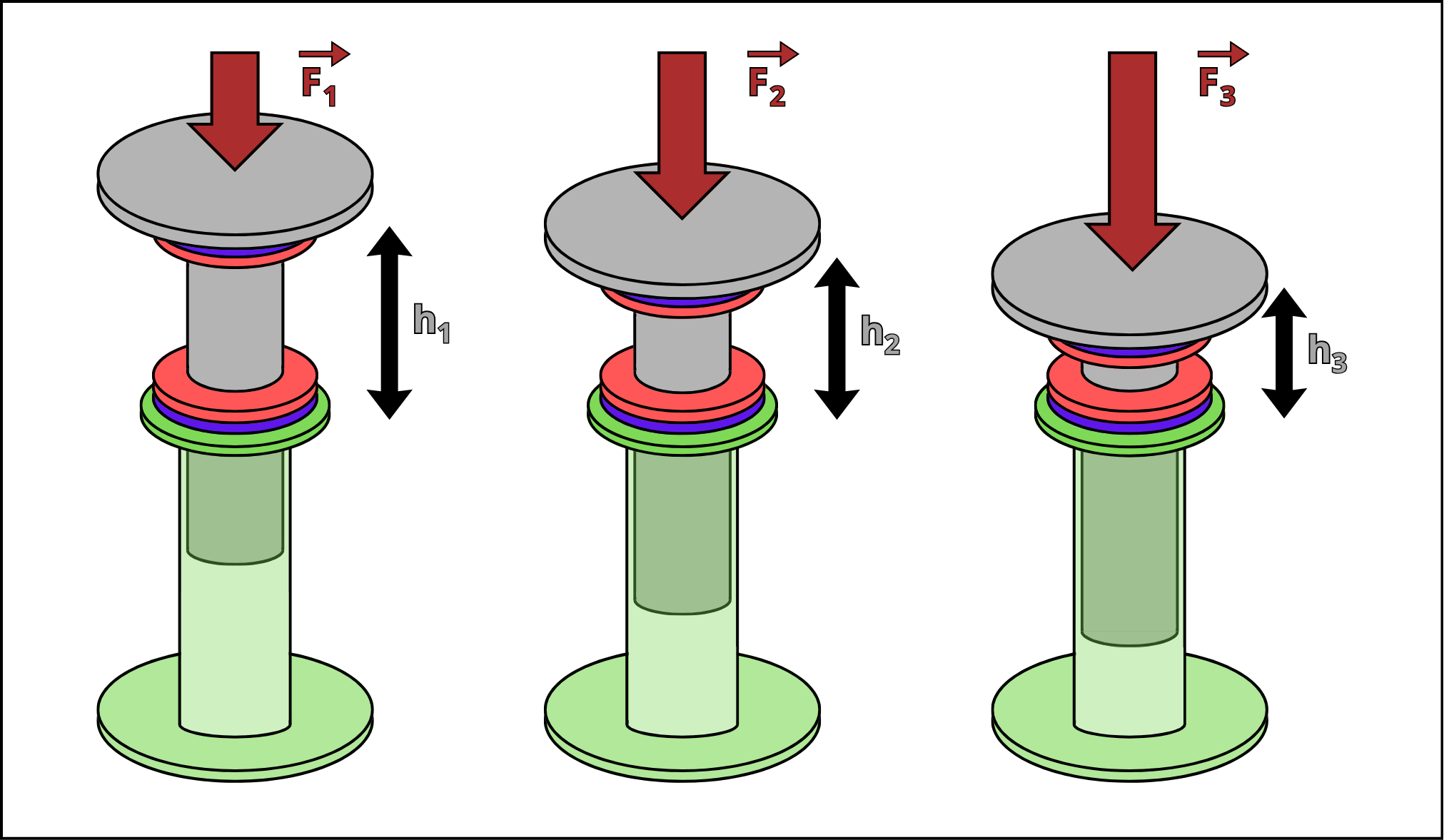}
\caption{Illustration of the experimental assembly. The base--piston device is shown (colored gray and green, respectively), with the two magnets facing each other so that the magnetic force is repulsive and the upper magnet levitates. Three stages are shown that differ in the mechanical load applied to the top of the device, reducing the separation $h$ between the magnets.}
\label{fig:montaje_experimental}
\end{figure}

The measurement protocol follows an incremental loading sequence, sketched in Fig.~\ref{fig:montaje_experimental}. 

The general procedure consists of the following steps: (i)~the mass of the levitating system---initially consisting of the upper magnet and the piston part---is determined to be $(28.740 \pm 0.005)~\text{g}$; (ii)~the setup is assembled as in Fig.~\ref{fig:montaje_experimental}, in a base--piston arrangement with the magnets facing each other so that the resulting magnetic force is repulsive; (iii)~the equilibrium separation $h$ is measured with a ruler (resolution of $0.5~\text{mm}$); and (iv)~known masses, consisting of a set of identical non-magnetic calibration weights with an individual mass of $m = (16.240 \pm 0.005)~\text{g}$, are sequentially added on the upper platform of the piston part, and the new equilibrium separation $h$ is recorded for each added mass. In this way we obtain a table of equilibrium separations $h$ as a function of the additional mass $m_{\text{add}}$, from which the magnitude of the magnetic force $F_{\text{mag}}$ can be computed.

\begin{figure}[h!]
    \centering
    \includegraphics[width=1\linewidth]{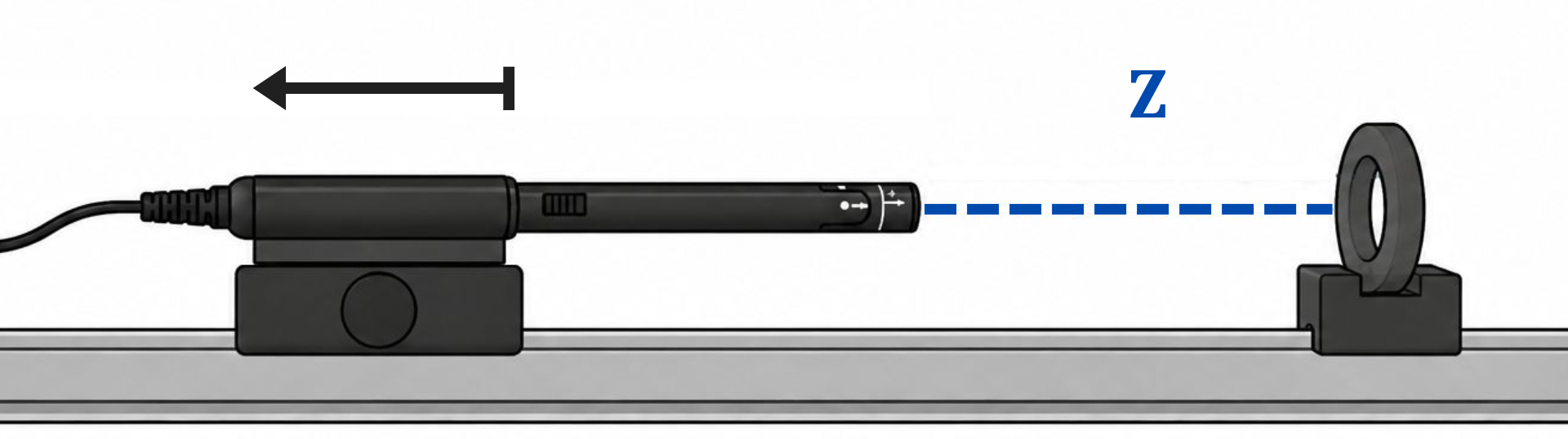}
    \caption{Schematic representation of the setup implemented during the measurements of $B(z)$ to ensure precise axial alignment on a rail track.}
    \label{fig:B_z}
\end{figure}

To complement the mechanical equilibrium analysis and provide a secondary, independent path for determining the magnetic moment $\mu$, a profiling experiment of the axial magnetic field $B(z)$ was carried out. A critical challenge and source of experimental limitation in this procedure is achieving a precise and stable concentric alignment between the active area of the magnetic probe and the main symmetry axis of the magnet. Slight lateral shifts or angular tilts can distort the measured field values significantly. To overcome this issue and ensure a highly controlled trajectory, a Vernier Hall-effect longitudinal probe (with an instrumental resolution of $0.2\text{ mT}$) and the target ring magnet were placed on a rigid, precision-machined rail track as depicted in Fig.~\ref{fig:B_z}. The magnet holder is held fixed at one end, while the Hall sensor mount slides smoothly along the rail. Measurements of the magnetic field intensity $B$ were systematically recorded at discrete intervals of $1$~cm, with the probe-tip reading $z$ ranging from $1$ to $20$~cm.

\section{Results and discussion}\label{sec:results}
\begin{figure}[h!]
\centering
\includegraphics[width=0.48\textwidth]{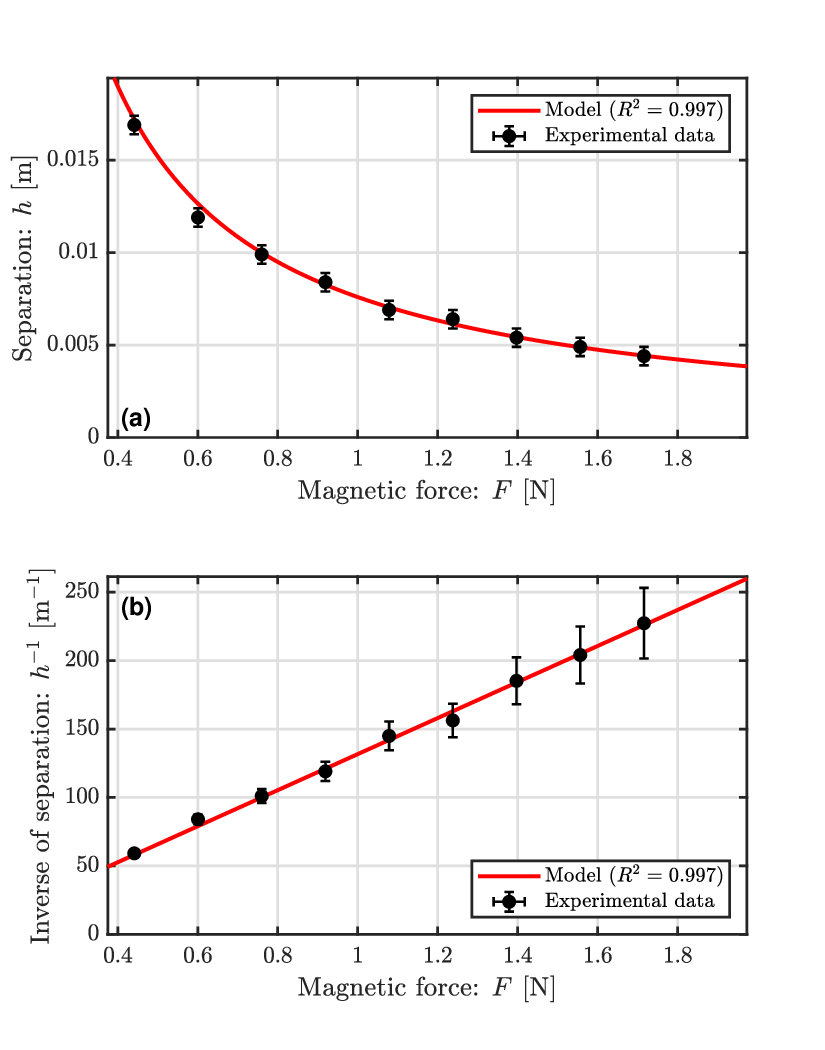}
\caption{Experimental results for the magnetic levitation system as a function of the applied load. (a)~Equilibrium separation $h$ as a function of the magnitude of the magnetic force $F_{\text{mag}}$. (b)~Linearization of the model, showing the inverse separation $h^{-1}$ as a function of the force magnitude $F_{\text{mag}}$. In both panels, the black markers represent the experimental measurements with their respective two-dimensional error bars, while the solid red line shows the prediction of the analytical model of Eq.~\eqref{eq:fuerza}.}
\label{fig:grafico_resultados}
\end{figure}

Figure~\ref{fig:grafico_resultados} shows the behavior of the separation $h$ as a function of the magnitude of the magnetic force $F_{\text{mag}}$. The experimental data were obtained by varying the total levitating mass and recording the equilibrium separation between the magnets; the mass was then converted to force magnitude units through Eq.~\eqref{eq:equilibrio}.

Figure~\ref{fig:grafico_resultados}(a) shows the nonlinear behavior of the separation as a function of the magnetic force magnitude. When the load on the upper device is small, slight variations in the force magnitude produce macroscopic changes in the levitation height. As the mass increases, however, the system requires substantially larger force increments to compress the separation between the magnets, tangibly revealing the nonlinear nature of the magnetic interaction.\cite{gonzalez_ejp_2017, forringer_tpt, robertson_ieee}

To subject the equivalent current-loop model to a strict quantitative test, Fig.~\ref{fig:grafico_resultados}(b) shows the linearized form of the system, obtained by plotting the inverse separation $h^{-1}$ against the magnitude of the magnetic force $F_{\text{mag}}$. A least-squares linear fit yields a coefficient of determination $R^{2} = 0.997$, validating the hypothesis of inverse proportionality between the magnetic force magnitude and the equilibrium separation.

Denoting the slope of Fig.~\ref{fig:grafico_resultados}(b) by $\beta$ (experimentally $\beta=(131.7\pm1.0)\,(\text{N}\,\text{m})^{-1}$), the magnetic dipole moment $\mu$ of the magnet can be directly determined by combining Eq.~\eqref{eq:fuerza} with the definition of the dipole moment for a planar current-loop, $\mu = I A$ (where $A = \pi R^2$ represents the area enclosed by the loop). This substitution leads to
\begin{equation}
\mu = \sqrt{\frac{\pi^2 R^3}{\beta\,\ln(4)\,\mu_{0}}}.
\label{eq:momento_despejado}
\end{equation}
We take as effective radius the outer radius of the rings, $R = 16$~mm. Applying Eq.~\eqref{eq:momento_despejado} to the slope above gives $\mu_{\text{lev}} = (0.420 \pm 0.002)~\text{Am}^2$, where the uncertainty comes only from the fit of $\beta$.

To independently evaluate the physical parameters of the permanent magnet, the on-axis magnetic field data $B(z)$ gathered via the rail track configuration were processed. Using the experimental values of $z$ and $B$, a least-squares fit was performed using equation \eqref{eq:Bz_model} to determine the magnetic dipole moment $\mu$ of the magnet (the loop radius was fixed at $R=16$~mm).

\begin{figure}[h!]
    \centering
    \includegraphics[width=0.94\linewidth]{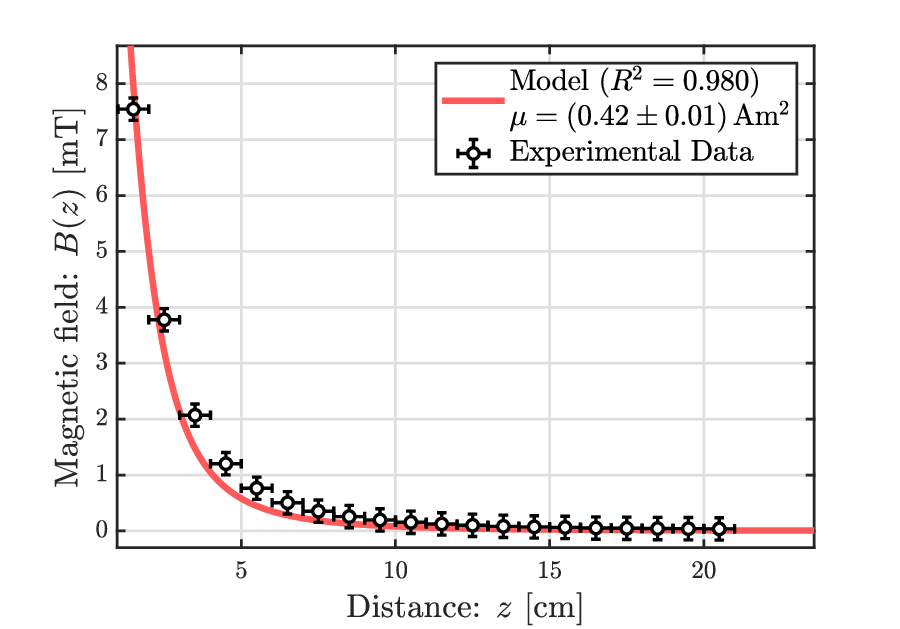}
    \caption{Experimental magnetic field $B(z)$ as a function of the axial distance $z$ measured with the Vernier Hall probe on the rail track. The black circular markers represent experimental data points with error bars, and the solid dark-red curve represents the equivalent current-loop model fit ($\mu = (0.420 \pm 0.012)~\text{A}\text{m}^2$).}
    \label{fig:Figura_B_z}
\end{figure}

The mapping of the magnetic field strength and its analytical fit are shown in Fig.\ref{fig:Figura_B_z}. Both independent techniques yield identical central values for the magnetic dipole moment, with the levitation equilibrium producing a higher-precision estimate of $\mu_{\text{lev}}=(0.420\pm0.002)\,\mathrm{A\,m^2}$ ($[0.418,0.422]\,\mathrm{A\,m^2}$), which lies entirely within the uncertainty interval of the field-fit determination ($[0.408,0.432]\,\mathrm{A\,m^2}$). This statistical consistency demonstrates that both configurations provide convergent characterizations of permanent magnetic materials in undergraduate laboratory frameworks, notwithstanding potential experimental nuances such as mechanical friction in the 3D-printed guide or sensor alignment precision.

Finally, it is worth acknowledging that modeling a thick ring magnet with a central hole as an ideal thin current-loop has inherent theoretical limitations. As demonstrated in more rigorous finite-element simulations, the $\vec{B}$-field along the axis of a ring magnet can exhibit non-monotonic behavior and even spatial reversal near the hole surface, phenomena which a simple loop model does not capture \cite{gonzalez_ejp_2017}. Nevertheless, our macroscopic measurements show that the equivalent loop approximation provides an adequate functional fit within the specific operational distances of our levitation setup, serving as an effective pedagogical tool before introducing advanced numerical methods.

\section{Conclusions}
In this work we have designed, built, and validated an instructional device for the quantitative study of magnetic levitation and the characterization of the magnetic forces between ring magnets. Through the strategic use of 3D printing, we circumvented the instability imposed by Earnshaw's theorem in an inexpensive, safe, and reproducible way, well suited to educational settings with limited resources.

Experimental results demonstrate that, even with the physical simplifications inherent in modeling real ferrite magnets as ideal current-loops, the proposed analytical model exhibits substantial predictive accuracy in the investigated laboratory range. Furthermore, an independent secondary methodology is introduced that uses a rail track to map the axial magnetic field $B(z)$. The central values of both independent methods coincide, and the higher-precision levitation result lies entirely within the uncertainty interval of the field mapping profile. This full statistical consistency strongly supports the validity of the equivalent current-loop model and confirms the robustness of the proposed approach.

From a pedagogical standpoint, the activity guides students effectively through a complete cycle of scientific inquiry: from mathematical modeling and instrument handling, through the methodical acquisition of data in mechanical equilibrium, to the estimation of microscopic properties of matter.
We therefore believe that this experiment turns a captivating phenomenon into a rewarding quantitative investigation. As frequently occurs with Olympiad problems, this low-cost, open, and ingenious proposal can be directly transferred to instructional laboratories, making it a valuable addition to the undergraduate curriculum.

\section{Availability of 3D design files}
\label{sec:availability}
The STL files used in this work are publicly available at Printables (\url{https://www.printables.com/model/1768224-magnetic-levitation-physics-experiment-educational}) and GitHub (\url{https://github.com/aronkahrs-us/Levitacion-Magnetica}).

\section*{Data availability statement}
The data that support the findings of this study are available within the article and the full numerical datasets are hosted in the associated GitHub repository. The 3D design (STL) files are openly available in the repositories indicated in Sec.~\ref{sec:availability}.

\section*{Acknowledgments}
The authors express their gratitude to the physics teaching staff at the Facultad de Ciencias, Universidad de la República (Montevideo, Uruguay), for their sustained commitment to developing innovative experimental undergraduate physics courses. In particular, we kindly acknowledge Dr.~M.~Reisenberger, whose creative lab problems for the \textit{Laboratorio de Física II} (2025) course inspired the design and development of the experimental setup presented in this work \cite{Reisenberger2025repartido3}.

\appendix
\section{Derivation of the Near-Field Approximation}
\label{app:deduction}

To justify the simplified force--distance relationship used in Eq.~\eqref{eq:fuerza}, we model the macroscopic ring magnets as equivalent Amperian surface-current cylinders (short solenoids) of finite axial thickness $w$ and radius $R$. In the experimental regime where the levitation height $h$ (the gap between the facing surfaces) is significantly smaller than the radius ($h \ll R$), curvature effects become negligible. The system can therefore be accurately approximated by the forces acting between two parallel current strips of width $w$ separated by a distance $h$.

The differential force per unit length between two infinite parallel line currents separated by a distance $y$ is governed by Amp\`ere's force law:
{\small
\begin{equation}
\frac{d(dF)}{d\ell} = \frac{\mu_0 \, dI_1 \, dI_2}{2\pi y} .
\end{equation}}

Assuming that the equivalent Amperian current $I$ is uniformly distributed across the thickness $w$ of each magnet, the current elements can be written as $dI_1 = (I/w)\,dz_1$ and $dI_2 = (I/w)\,dz_2$. To define a clear coordinate system with explicit integration limits, let the lower magnet occupy the region $z_1 \in [-w, 0]$ and the upper magnet occupy $z_2 \in [h, h+w]$, with the origin $z=0$ fixed at the top surface of the lower magnet. The geometric distance between any two differential elements is strictly $y = z_2 - z_1$, which remains positive throughout the domain. Integrating over the thickness of both magnets yields the total force per unit length:
{\small
\begin{equation}
\frac{dF}{d\ell} = \left[\frac{\mu_0 I^2}{w} \right]\left(\frac{1}{2\pi w} \int_{h}^{h+w} \int_{-w}^{0} \frac{1}{z_2 - z_1} \, dz_1 \, dz_2 \right) .
\end{equation}}

Evaluating the first integral with respect to $z_1$ gives:
{\small
\begin{align*}
\int_{-w}^{0} \frac{1}{z_2 - z_1} \, dz_1 &= \left[ -\ln(z_2 - z_1) \right]_{-w}^0 \\
&= \ln\left( \frac{z_2 + w}{z_2} \right) .
\end{align*}}

Subsequent integration of the remaining logarithmic terms with respect to $z_2$, from the lower limit $h$ to the upper limit $h+w$, results in the following exact geometric expression:
{\small
\begin{align*}
\frac{dF}{d\ell}
&=  \left[\frac{\mu_0 I^2}{w} \right]\left(\frac{1}{2\pi w}\right)  [(h+2w)\ln(h+2w) \\
& \quad\quad\quad\quad\quad\quad\quad\quad- 2(h+w)\ln(h+w) + h\ln(h)] .
\end{align*}}

In the extreme near-field approximation, where the surfaces are nearly in contact ($h \ll w$), the logarithmic terms admit a simple asymptotic limit. Since $\lim_{h\to 0} h\ln(h) = 0$, evaluating the remaining terms in the contact limit reduces the bracketed expression directly to $[\,w\ln(4)\,]$, that is,
{\small
\begin{align*}
    \frac{dF}{d\ell}  &=\lim_{h\to0}  \left[\frac{\mu_0 I^2}{w} \right]\left(\frac{1}{2\pi w}\right)  [(h+2w)\ln(h+2w) \\
& \quad\quad\quad\quad\quad\quad\quad\quad- 2(h+w)\ln(h+w) + h\ln(h)] \\
& = \left[\frac{\mu_0 I^2}{w} \right]\left(\frac{1}{2\pi w}\right) [2w \ln(2w)-2w\ln(w)] \\
& = \left[\frac{\mu_0 I^2}{w} \right]\left(\frac{1}{2\pi w}\right) [w \ln(4)] .
\end{align*}}

To compute the total macroscopic magnetic force $F_{\text{mag}}$, we multiply the force per unit length by the effective mean perimeter of the ring magnet, $2\pi R$:
{\small
\begin{align}
F_{\text{mag}} &= \left( \frac{dF}{d\ell} \right) \cdot (2\pi R) = \ln(4)\frac{\mu_0 I^2 R}{w} .
\end{align}}

This result shows that, under extreme near-field conditions, the volumetric interaction scales inversely with the magnet thickness $w$. Physically, this behavior can be mapped onto an effective model of concentrated thin loops. When the distributed Amperian currents are treated as concentrated filaments localized at the interaction boundary, the operational length scale governing the field gradient collapses onto the inter-surface separation $h$. Substituting $w$ by this effective separation length scale ($h$) yields the final pedagogical force law presented in Eq.~\eqref{eq:fuerza}. While this substitution relies on the heuristic physical assumption that the field gradient in the near-field is overwhelmingly dominated by the gap distance $h$, it yields a surprisingly accurate and practical macroscopic functional form for the force scaling in this operational regime \cite{saslow_tpt_2022}.


\begin{thebibliography}{99}

\bibitem{arribas2015measurement}
E. Arribas, I. Escobar, C. P. Su\'arez, A. N\'ajera, and A. Bel\'endez,
``Measurement of the magnetic field of small magnets with a smartphone: A very economical laboratory practice for introductory physics courses,''
\textit{Eur. J. Phys.} \textbf{36}(6), 065002 (2015).

\bibitem{gonzalez_ejp_2017}
M. I. Gonz\'alez,
``Forces between permanent magnets: Experiments and model,''
\textit{Eur. J. Phys.} \textbf{38}(2), 025202 (2017).

\bibitem{forringer_tpt}
E. R. Forringer,
``Measuring and modeling the force between permanent magnets,''
\textit{Phys. Teach.} \textbf{60}(7), 546--548 (2022).

\bibitem{tada_tpt_2026}
K. Tada,
``Modular Tracks for Magnetic Levitation,''
\textit{Phys. Teach.} \textbf{64}(5), 362--363 (2026).

\bibitem{earnshaw}
S. Earnshaw,
``On the nature of the molecular forces which regulate the constitution of the luminiferous ether,''
\textit{Trans. Camb. Phil. Soc.} \textbf{7}, 97 (1848).

\bibitem{scott_earnshaw}
W. T. Scott,
``Who was Earnshaw?,''
\textit{Am. J. Phys.} \textbf{27}(6), 418--419 (1959).

\bibitem{rossing_hull_1991}
T. D. Rossing and J. R. Hull,
``Magnetic levitation,''
\textit{Phys. Teach.} \textbf{29}(9), 552--562 (1991).

\bibitem{simon_ajp_1997}
M. D. Simon, L. O. Heflinger, and S. L. Ridgway,
``Spin stabilized magnetic levitation,''
\textit{Am. J. Phys.} \textbf{65}(4), 286--292 (1997).

\bibitem{berry_geim_1997}
M. V. Berry and A. K. Geim,
``Of flying frogs and levitrons,''
\textit{Eur. J. Phys.} \textbf{18}(4), 307--313 (1997).

\bibitem{osorio_2012}
M. R. Osorio, D. E. Lahera, and H. Suderow,
``Magnetic levitation on a type-I superconductor as a practical demonstration experiment for students,''
\textit{Eur. J. Phys.} \textbf{33}(5), 1383--1395 (2012).

\bibitem{giliberti_ejp_2018}
M. Giliberti, L. Perotti, and L. Rossi,
``Motion of a superconducting loop in an inhomogeneous magnetic field: A didactic experiment,''
\textit{Eur. J. Phys.} \textbf{39}(5), 055503 (2018).

\bibitem{meissner_1933}
W. Meissner and R. Ochsenfeld,
``Ein neuer Effekt bei Eintritt der Supraleitf\"ahigkeit,''
\textit{Naturwissenschaften} \textbf{21}(44), 787--788 (1933).

\bibitem{strehlow_ajp_2009}
C. P. Strehlow and M. C. Sullivan,
``A classroom demonstration of levitation and suspension of a superconductor over a magnetic track,''
\textit{Am. J. Phys.} \textbf{77}(9), 847--851 (2009).

\bibitem{tomes_ejp_2016}
J. J. Tomes and C. E. Finlayson,
``Low cost 3D-printing used in an undergraduate project: An integrating sphere for measurement of photoluminescence quantum yield,''
\textit{Eur. J. Phys.} \textbf{37}(5), 055501 (2016).

\bibitem{bley_ejp_2021}
J. Bley, A. Pietz, A. F\"osel, M. Schmiedeberg, S. Heusler, and A. Pusch,
``Physics competitions in the time of a pandemic: 3D printing as a new approach to the quantitative investigation of Cartesian divers at home,''
\textit{Eur. J. Phys.} \textbf{43}(1), 014001 (2022).

\bibitem{su_ajp_2016}
J. Su, W. Wang, M. Lu, X. Xu, Q. F. Yan, and J. Lu,
``Visualization of gravitational potential wells using 3D printing technology,''
\textit{Am. J. Phys.} \textbf{84}(12), 943--947 (2016).

\bibitem{rossi_ajp_2021}
E. Rossi,
``The Newtonian gravity of irregular shapes using STL files and 3D printing,''
\textit{Am. J. Phys.} \textbf{89}(11), 993--1001 (2021).

\bibitem{perez_ajp_2019}
A. T. P\'erez, P. Garc\'ia-S\'anchez, M. A. S. Quintanilla, and A. Fern\'andez-Prieto,
``Levitation? Yes, it is possible!,''
\textit{Am. J. Phys.} \textbf{87}(4), 270--274 (2019).

\bibitem{malmstrom_ejp_2020}
H. Malmstr\"om, J. Enger, M. Karlsteen, and J. Weidow,
``Integrating CAD, 3D-printing technology and oral communication to enhance students' physics understanding and disciplinary literacy,''
\textit{Eur. J. Phys.} \textbf{41}(6), 065708 (2020).

\bibitem{vuckovic_icest}
A. N. Vu\v{c}kovi\'c, S. S. Ili\'c, and S. R. Aleksi\'c,
``Calculation of the attraction force between a permanent magnet and an infinite linear magnetic plane using Ampere's currents,''
in \textit{Proc. ICEST 2011}, 489--492 (2011).

\bibitem{saslow_tpt_2022}
W. M. Saslow,
``Magnetic Poles: A Missing Manual,''
\textit{Phys. Teach.} \textbf{60}(7), 540--545 (2022).

\bibitem{robertson_ieee}
W. S. P. Robertson, B. Cazzolato, and A. C. Zander,
``A simplified force equation for coaxial cylindrical magnets and thin coils,''
\textit{IEEE Trans. Magn.} \textbf{47}(8), 2045--2049 (2011).

\bibitem{ref_1_13365}
S. Defrancesco and V. Zanetti, ``Experiments on magnetic repulsion,'' \textit{Am. J. Phys.} \textbf{51}(11), 1023--1025 (1983).

\bibitem{ref_2768690}
L. E. Gayetsky and C. L. Caylor, ``Measuring the forces between magnetic dipoles,'' \textit{Phys. Teach.} \textbf{45}(6), 348--351 (2007).

\bibitem{ref_0143_0807}
K. Li, J. Lin, Z. Y. Kang, S. Y. W. Liang, and J. Wong Say Juan, ``The vertical oscillations of coupled magnets,'' \textit{Eur. J. Phys.} \textbf{32}(4), S1 (2011).

\bibitem{griffiths_ed}
D. J. Griffiths, \textit{Introduction to Electrodynamics}, 4th ed. (Pearson, Boston, 2013), Example 5.6.

\bibitem{Reisenberger2025repartido3}
M.~ Reisenberger, \textit{Repartido N°3 Electromagnetismo}, Material didáctico para el curso \textit{BQ424-Laboratorio de Física II para Bioquímica}, Instituto de Física, Facultad de Ciencias, Universidad de la República, Montevideo, 2025.
\end{thebibliography}
\end{document}